# We Need to Rethink How We Describe and Organize Spatial Information

Instrumenting and Observing the Community of Users to Improve Data Description and Discovery

Benjamin Adams and Mark Gahegan
Centre for eResearch and Dept. of Computer Science
The University of Auckland
Auckland, New Zealand
e-mail: {b.adams,m.gahegan}@auckland.ac.nz

*Abstract*—In Spatial Data Infrastructure or Cyber Infrastructure, the description of geographic data semantics is intended to support data discovery, reuse and integration. In the vast majority of cases the producers of these data generate descriptions based on particular understandings of what uses the data are good for. This producer-oriented perspective means that the descriptions often do not help to answer the question of whether a data set is of use for a consumer who might want to apply it in a different context. In this paper, we discuss the role geographic information observatories can play in providing an infrastructure for observing the context of data use by consumers. These observations of data pragmatics lead to operational statistical methods that will support better fitness-for-use assessment. Finally, we highlight some of the challenges to building these observatories, and briefly discuss strategies to address those challenges.

I. INTRODUCTION

The goal of fostering data reuse and semantic integration by describing geographic data in a cyberinfrastructure has met with limited practical success, despite many years of research. One of the challenges that such efforts face is the *situated* nature of geographic knowledge—how we understand data depends strongly on our own experience and expertise, and potentially also the situation within which we intend to use it [7]. Systems that organize information are usually designed to support a set of interactions by a community of users [8]. But, if the organizational system that is used in a cyberinfrastructure does not incorporate a model of the user community—and react to what can be learned from user interactions with the infrastructure—then it will have limited utility. To date the onus for describing data has primarily fallen on the producers of the data, whether they are authoritative organizations or individual scientists. This approach means that the descriptions usually reflect what the *producers* value in the data, without consideration of whether the data will be fit-for-use by a potential *consumer* of the data. This is entirely reasonable given that the producer cannot possibly anticipate all the ways in which their data might be used. But perhaps we are approaching the problem incorrectly in thinking of it this way? To further complicate matters, the heterogeneity of geographic data, and the differing goals of its producers has lead to data being described in a multitude of incompatible ways. And despite a large number of research papers on geospatial semantics—are often not getting any easier to reconcile in practice [15]. The result is a confusion of *just-so data stories*, describing what data "means", but in a manner that provides very little help for data consumers to find the data that is suitable for their purposes.

This is not a new critique. Frank [5] identified this problem (phrased in terms of data quality) well before the era of using semantic web languages to describe geographic data. But the critiques of producer-oriented data description (see also [3][17]), have not led to improved operational approaches for organizing geospatial information, perhaps because their strategies are quite abstract in their own way, lacking a clear methodology for putting them into practice. Current efforts on geo-semantics, in contrast, make operationalization of languages (semantics) and reasoning paramount, without much consideration of fitness-for-use [11]. If, as has been argued, fitness-for-use assessment is critical for functional Spatial Data Infrastructure (SDI) (or Geo-CyberInfrastructure), then we need practical and achievable methods that allow us to understand the context of data use from the perspective of the user: that is, in contexts that may not have been foreseen by the data producers or cyberinfrastructure builders.

In this paper, we explore how some of the emerging ideas from *information observatories* could be applied in a geographical context to provide a practical solution for including the consumer in our methods of data description [1]. The approach we advocate is not without its own challenges (many of them social), but if adopted by the community, we believe will lead to statistical methods that will allow us to better support fit-for-use data assessment in cyberinfrastructure.

An Information Observatory is infrastructure designed for understanding the ecosystem of information that *observes* not just the object of study but also the conceptual structures, data, and actors involved in the process of analysis and knowledge production, from multiple perspectives. A Geographic Information Observatory (GIO) is thus an information observatory focused on geographic information AND its community of practice. Building GI observatories means building infrastructure that can observe geospatial data in the context of its use, within a community. This is quite distinct from traditional approaches to data description in cyberinfrastructure, which focus primarily on using metadata to describe data formats and the semantics of data content: in an information observatory we consider not only data but also tasks and methods performed with the data, the domain knowledge of data producers and consumers, communities-of-practice, and more; all of these facets become first-class observable artifacts (signifiers in the semiotic sense) that carry meaning and help to explain or contextualize each other.

We propose that GI Observatories can provide insight into the dynamic, geographical scientific process from multiple perspectives, from data through tasks and methods to communities-of-practice, and that this insight will help us build cyberinfrastructure that will better support these interactions [1][6]. Or to put it another way, we propose to make an empirical science out of cyberinfrastructure development. We want to observe the relationships between these different facets of geographic information, because these observables are only meaningful when brought into relation with other concepts. And by observing them, we can learn from them all kinds of practical insights that can help characterize what information is used for, by whom, using which methods, for what tasks. Over time, such observations can yield actionable intelligence that may add significant value over and above what can be achieved by formal semantics.

In the following section we review related work. In Section III we discuss the nexus of knowledge relations that are employed in geographic research. Section IV details how GI Observatories can be implemented by operationalizing the nexus as a graphical model. In Section V we discuss some considerations of the role of the community in building GI Observatories, and we conclude with a discussion of the opportunities and challenges of building these observatories going forward.

## II. RELATED WORK

The Web Observatory is a nascent idea proposed to support the notion of doing Web Science. That is, observing the web in order to understand how human activity shapes the Web, and how human activity patterns and the Web co-evolve. Web Observatories have been described as the "middle layer for broad data" meaning that, as data production has become more distributed and decentralized, the ability to perform analyses on these data has remained siloed, and the observatory serves to open up that analytic framework [19]. To date, the development of web science observatories has focused mostly on data collection / mashup tools that produce views on 'big' web data such as streaming social media content [20][21]. Recent work exploring the idea of building observatories on top of citizen science projects, such as *Zooniverse*, point to promising applications of Web Observatories to the sciences [22].

Although there are some superficial similarities between Web Observatories and the GI Observatory idea that we are proposing with respect to observing human information interaction, our focus differs in two important ways. First, the Web Observatory is a macro-scale observatory in the sense that it is designed to provide analytic infrastructure to explore properties of the web and human society, whereas the Information Observatory aims to capture observations at the granularity of data use and change, by individual researchers and within specific scientific communities [23]. Second, Web Observatories are primarily described as tools for analysis of the Web as a socio-technical system, thus there is a very specific subject of analysis, namely the Web. As noted, our motivation for building GI Observatories is in large part driven by a desire to build better cyberinfrastructure for scientific discovery, and we are interested in understanding the universe of information from a multitude of perspectives.

Personalization in information retrieval requires modeling the background context under which a user performs an information-searching task [24][25]. This context model can take the form of an explicit user model or can be based on other kinds of implicit behavioral feedback, such as search history and click-through data [26][27][28]. Group level models of personalization are based on the notion that similar users will want similar search results [29]. Recommender systems built with collaborative filtering algorithms fall within this category of personalization [30].

Several variables can play a role in creating a user model using a relevance feedback framework. For example, the temporal scope of the information can be important, so that immediate search history might be more relevant than longer-term behavior [31]. Beyond search history, measures of user interests and activities from heterogeneous sources (documents, emails, etc.) can also be useful [32]. In social search, the role of the user within a larger community becomes an important factor [33].

Personalized search based on past behavior has raised concern about the potential drawback of creating a filter bubble, where potentially relevant information is not shown because of the personalization algorithm [34]. The many personalization methodologies and algorithms that have been developed for information retrieval and web search could well be applied to observational data collected by a GI Observatory and then be used to develop new ways of searching for scientific data.

## III. THE NEXUS OF RELATIONS

Inspired by Alfred North Whitehead's [18] writings on the intricate web of relationships that participates in knowledge representation, Gahegan and Pike [7] described a *nexus* of relations that link the many conceptual structures used in geographic research. Their nexus is shown in Figure 1. Based on this nexus the Codex system was built to capture these relations and make it possible for a user to explore resources through any and all of these relations. In practice, however, depending on the content of each of the nodes in the nexus and context that is of interest to the user, only a small subset of those relations will actually be relevant. What is missing is a way to structure the web of relations in a way that facilitates understanding without overloading the user with unnecessary information.

The ovals in Figure 1 represent the conceptual structures that the GIScience research community has expended most effort on describing, in some cases building metadata standards for describing those concepts. Semantic metadata descriptions of geographic data in cyberinfrastructure focus on "objective" aspects of geographic data shown as circles in blue. However, the nodes (purple clouds) might take on very different values depending on whether one is a producer or consumer of data (the researcher in red). The purple nodes—which are usually not captured—are important to assess fitness-for-use.

For example, we have semantic models of measurement, geographic data models, and scientific workflows [4][9][13]. Furthermore, current GIScience usually has its gaze fixed on Geographic Information (as its name suggests), so this is the subject around which other concepts are positioned. Semiotically, we could say that information (or data) is always the interpretant, with other data used to help explain it. But this misses an opportunity to focus on other key facets, such as a researcher or a method and to use data to help describe them. Meanwhile, situational knowledge is not formally captured (represented by the cloud shapes in Figure 1), other than in some cases natural language, e.g., in journal publications. Where other facets are captured, it is usually referred to as *context*, but now let's recognize that it is simply a set of observations onto a different set of objects, not typically represented in an infrastructure, but could be.

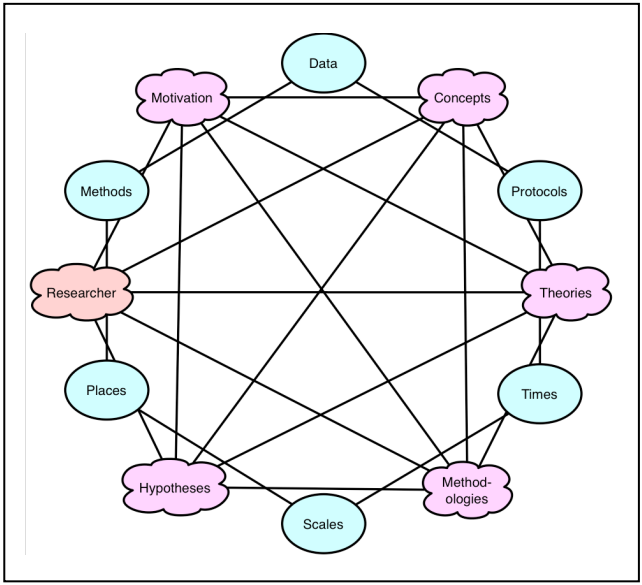

Figure 1. The nexus of relations between (some of) the conceptual structures used in geographic research, adapted from Gahegan & Pike [7].

Arguably, scientists who use these concepts in their work are more likely to be able to articulate what they want and what is important to them in terms of the conceptual structures represented by the cloud shapes—motivations, methodologies, theories, etc., rather than using the formal description languages for data that have been the target of most work on geosemantics. This in no small part might explain why—despite great effort in GIScience to advance semantics—semantic technologies are not being used by most geo-scientists in their day-to-day research. It is simply not the language that they use to think about and communicate their research.

The model of the nexus is just one example meta-model for the kinds of relationships between conceptual knowledge used in geographic research. For example, a sub-graph of these relations forms the graphical model of data production described by Gahegan and Adams [7]. In that model, each node in the nexus (*community*, *task*, *domain knowledge*, and *data*) are described by facets intended to capture the *who*, *what*, and *why* of data. The relations in that model are also directed, rather than undirected as in the original nexus meta-model.

When building cyberinfrastructure for data re-use, we are faced with a two-fold problem: first, whether the representation that we choose for each node in the nexus is a functionally useful representation (such as an ontology of conceptual structures) for scientists engaging in geographic knowledge production; and second, whether the meta-model describing the relations and directionality between the nodes in the nexus is itself useful. The GI Observatory gives us an infrastructure for addressing the first problem by collecting data on how conceptual structures relate to one another in practice, and then evaluating the information to be gained about one facet from a particular description of another facet. The second problem is encapsulated in the dynamic nature of the GI Observatory and its community-driven mandate—they are built to serve geo-scientists and the meta-model should be flexible and revised as a richer and more accurate understanding of the community's needs emerges. In this sense, Figure 1 is just a place to start, not a final recommendation.

To get us beyond the essentially meaningless result that everything is related to everything else, a GI Observatory will always need to make a commitment to a given meta-model in order to observe the nodes in the nexus. We also need to identify where to fix the GI Observatory's "telescope" and where to point it. For example, we cannot look at *tasks* from a fixed perspective without deciding first what information artifacts of *tasks* act as signifiers and then describing the object of those signs, whether they be *data*, *methods*, *communities*, etc. Most importantly, many different kinds of meta-models and signifiers and objects of enquiry can be created and could even co-exist, and the choice of those meta-models will depend on what the users of the GI Observatory want to use it for.

IV. OPERATIONALIZING THE NEXUS AS A GRAPHICAL MODEL

Relational network representations of knowledge such as the one shown in Figure 1 are found in all kinds of information systems. And such graphical models can be recast as learning problems that are solved by doing statistical inference on networked nodes that represent variables that can take on many values [12]. Because mature computational patterns have been built for statistical inference on graphical models, they are applied to stochastic problems in a wide range of fields, including speech recognition, natural language processing, statistical physics, spatial statistics, and bioinformatics. The network of conceptual structures used in practice and observed through a GI Observatory can similarly be modeled as graphical models. The variables observed can span across data (formats, semantics, geometric models), tasks, analytical methods, computational workflows and people (researchers, communities).

Graphical webs of relations (Figure 1) can be either undirected or directed. A directed graph (mathematically represented as an acyclic graph) is used to model cause-and-effect relationships, whereas an undirected graph can be used when the causal structure is undefined. The model presented in [6] is directed and explicitly represents data production as a generative model with variables that describe causal relationships between communities that perform tasks with domain knowledge, which results in data of some kind. In contrast, the nexus in [7] is undirected, nodes are connected but there is no explicit directionality to the connections. Both kinds of models can form the backbone of a GI Observatory and utilizing the methodologies of statistical inference on graphical models (such as Bayesian Belief Network Learning or more general Inductive Model Discovery) can provide unique insights into the geographic information ecosystem [2][14]—and importantly without having to choose beforehand which causes what.

However, in order to access these patterns through data-driven discovery we need sufficient data, and it is in providing access to data about people (both producers and consumers of data), methods, workflows, and intention that GI Observatories show real promise in improving our geo-infrastructure. Several strategies can be employed to observe the variables in the nexus of relations. The lowest hanging fruit is to extract pragmatic relationships connecting people, ideas and things to datasets from natural language available in data repositories and on the web. Figure 2 shows how some of this information is encoded already in abstracts and webpages for researchers and investigating organizations. The text highlighted in purple in Figure 2 represents information that can be extracted from natural language and mapped to entities that represent motivations, concepts, theories, etc. In this example the data producer describes very specific affordances that the data provides, but we know very little about how the consumers of the data have used it and whether the entities that the producer highlighted are in fact of high informational value to potential consumers.

## V. INVOLVING A COMMUNITY OF USERS

GI Observatories will need to be built to serve the scientific community, so scientists should get real value out of what we can learn from these Observatories. Thus, ideally we involve geographers and domain scientists in their design, building, and re-building [1]. One of the most significant challenges to building GI Observatories is that, in order to observe many aspects of the scientific process, the Observatory will need access to information about what researchers do that is usually not readily available. For example, what kind of methods are scientists who are working in a specific field of research using with what kinds of data? Here we can learn from recent work by David Ribes on what he calls *scaling up ethnography*: "The object of analysis for the ethnographer ... becomes the methods, techniques and technologies used by actors to know and manage their enterprise." [16]. The instrumenting of the community to better understand data use presents several ethnographic challenges. But the fact that researchers increasingly interact with data and methods through infrastructures and workflow scripts means that at least some of the information we seek is in fact readily available. We simply need to instrument the SDIs to record it.

A GI Observatory provides the research platform for doing scaled up ethnography on not only large-scale geospatial cyberinfrastructure projects, such as spatial data infrastructures, instantiations of Digital Earth, and smart cities; but also the activities of the geographic scientific community writ large.

## VI. RESULTS AND FINDINGS

A prototype system for capturing the patterns of interaction between a community of users, tasks, methods, concepts and datasets has been created. It uses the notion of Description Spaces for: Space and Time, Domain Semantics, Processes and Community. Each space contains a smaller number of descriptive attributes. The Description Spaces allow us to compute a compound distance score between any pair of items, such as a researcher and a dataset or a method and a task. Over time, any use case that connects such items is remembered, effectively that can be used to 'bring them closer'. Bayesian inference is used on these Description Spaces and past histories to predict likelihood values that a certain researcher might be interested in a certain dataset or method, and so forth: in essence a recommender system but drawing from a much richer description of the problem domain than is usually found in conventional SDI or GIS. A complete account, with examples, is provided in [6].

In comparison to current SDI, this new approach simply broadens the focus, so everything we now consider as SDI still applies, but in addition we must (i) broaden the conceptual model used along the lines of the Descriptions Spaces describe above and (ii) 'instrument' the community of users so that we study and learn from what they do, thus it represents a much broader focus. But in return, these enhancements offer a more complete picture of how communities operate in practice, and we believe that such knowledge is extremely valuable, on a par with theory in terms of its usefulness to researchers. As an example: a recommendation such as this might be very helpful: "most climate change impacts researchers so far have used *this* interpolation method with *that* kind of dataset when working on coastal erosion problems".

Of course, these additional insights require additional effort to design and build the SDI initially and more research is yet needed to find: (i) ways to capture use patterns unobtrusively during the research process and (ii) ways to insert recommendations and insights back into the research process.

## VII. CONCLUSIONS AND FUTURE WORK

Easy discovery, reuse and integration of geospatial data have been the predominant motivations for decades of research on geosemantics. However, despite concerted effort by the GIScience community, a large gap continues to exist between the aspirations of the geosemantics research and applied outcomes that are used in the daily work of scientists. This stands in contrast to the very successful adoption of other GIScience research (e.g., spatial statistics), which has found wide adoption and successful implementations in a variety of geographic information systems.

We argue that the reason for this state of affairs is that the way working scientists assess the "meaning" and quality of geographic data is based on wide variety factors that are not currently brought together in a holistic way in our semantic technologies. The *situated context* of data use comprises more that just the "meaning of the data" from the perspective of the producer. It is also who has used it, how they have used it, and why they have used it. All of these properties of data use are potentially valuable pieces of information to aid discovery, reuse and integration.

The nexus of relations surrounding geographic data use represents a complex system composed of simpler interrelated parts—each of these simpler parts is possible to observe and measure. If we build our cyberinfrastructure to observe this wider context of data use, then we have an opportunity to build systems that can describe data in ways that better match how scientists themselves assess the fitness of data for their purposes. In many other fields, graphical and probabilistic methods that can learn structure in the face of complexity and uncertainty have demonstrated significant success in the last decade. We contend that if we are successful in instrumenting our community to observe this complex network of relations surrounding data use, then we too can put these powerful methods of inference to build cyberinfrastructure that better works with geoscientists, and finally begin to realize the long stated potential of geosemantic technologies.

Additional work is required to refine the Description Spaces used so that they are both useful (richly describe the problem domain) and easy to use (simple for the user to grasp). Descriptions used must be either easy for a user to enter, or easy to learn via use-cases. We make no claims that our current Description Spaces are optimal, merely that they are useful. But others could be more useful. If such Description Spaces are treated as meta-models, it should be possible to test different arrangements, to see which offer the best balance between usefulness and simplicity. As noted already, additional research is also needed to find ways to unobtrusively harvest useful information during the research process (for example from workflows) and to then make it available to future researchers as recommendations.

REFERENCES

[1] B. Adams, M. Gahegan, P. Gupta, and R. Hosking, "Geographic information observatories for supporting science." In Proceedings of Workshop on Geographic Information Observatories, 2014, pages 1-5.
[2] W. Bridewell, P. Langley, L. Todorovski, and S. Džeroski, "Inductive process modeling." Machine learning, vol. 71, no. 1, 2008, 1-32.
[3] R. Devillers, Y. Bédard, R. Jeansoulin, and B. Moulin, "Towards spatial data quality information analysis tools for experts assessing the fitness for use of spatial data." International Journal of Geographical Information Science, vol. 21, no. 3, 2007, 261-282.
[4] F. T. Fonseca and M. J. Egenhofer, "Ontology-driven geographic information systems." In Proceedings of the 7th ACM international symposium on Advances in geographic information systems, 1999, pages 14-19, ACM.
[5] A. U. Frank, "Metamodels for data quality description." In M. Duckham, E. Pebesma, K. Stewart, A.U. Frank editors, Data Quality in Geographic Information-From Error to Uncertainty, 1998, pages 142–158. Springer.
[6] M. Gahegan and B. Adams, "Re-envisioning data description using Peirce's pragmatics." In M. Duckham, E. Pebesma, K. Stewart, A.U. Frank editors, Geographic Information Science, Lecture Notes in Computer Science, vol. 8728, 2014, pages 142–158. Springer.
[7] M. Gahegan and W. Pike, "A situated knowledge representation of geographical information." Transactions in GIS vol. 10, no. 5, 2006, 727–749.
[8] R. Glushko, "Foundations for Organizing Systems." The Discipline of Organizing, 2013, pages 1-45, MIT Press.
[9] C. Granell, R. Lemmens, M. Gould, A. Wytzisk, R. de By, and P. van Oosterom, "Integrating semantic and syntactic descriptions to chain geographic services." Internet Computing, IEEE, vol. 10, no. 5, 2006, 42-52.
[10] K. Janowicz, B. Adams, G. McKenzie, and T. Kauppinen, "Towards Geographic Information Observatories." In Proceedings of Workshop on Geographic Information Observatories, 2014, pages 1-5.
[11] K. Janowicz, S. Scheider, and B. Adams, "A geo-semantics flyby." In Reasoning Web. Semantic Technologies for Intelligent Data Access, 2013, pages 230-250, Springer.
[12] M. I. Jordan, "Graphical models." Statistical Science vol. 19, no. 1, 2004, 140–155.
[13] W. Kuhn, "A functional ontology of observation and measurement." In GeoSpatial Semantics, 2009, pages 26-43, Springer.
[14] W. Lam and F. Bacchus, "Learning Bayesian belief networks: An approach based on the MDL principle." Computational intelligence, vol. 10, no. 3, 1994, 269-293.
[15] C. L. Palmer, M. H. Cragin, P. B. Heidorn and L. C. Smith, "Data curation for the long tail of science: The case of environmental sciences." In Third International Digital Curation Conference, Washington, DC, 2007, pages 1-5.
[16] D. Ribes, "Ethnography of scaling, or, how to a fit a national research infrastructure in the room." In Proceedings of the 17th ACM Conference on Computer Supported Cooperative Work & Social Computing, CSCW '14, 2014, pages 158–170. ACM, New York, NY, USA.
[17] R. Y. Wang and D. M. Strong, "Beyond accuracy: What data quality means to data consumers." Journal of management information systems, vol. 12, no. 4, 1996, 5-33.
[18] A. N. Whitehead, Process and Reality: An Essay in Cosmology. New York, Social Science Book Store, 1929.
[19] T. Tiropanis, W. Hall, J. Hendler, and C. de Larrinaga, "The web observatory: A middle layer for broad data." Big Data, vol. 2, no. 3, 2014, 129–133.
[20] K. McKelvey and F. Menczer, "Design and prototyping of a social media observatory." In Proceedings of the 22nd international conference on World Wide Web companion, 2013, pages 1351–1358. International World Wide Web Conferences Steering Committee.
[21] X. Gao, et al., "Supporting a social media observatory with customizable index structures: Architecture and performance." In Cloud Computing for Data-Intensive Applications, 2014, pages 401–427. Springer.

[22] R. Simpson, K. R. Page, and D. De Roure, "Zooniverse: observing the world's largest citizen science platform." In Proceedings of the companion publication of the 23rd international conference on World wide web companion, 2014, pages 1049–1054. International World Wide Web Conferences Steering Committee.

[23] K. O'Hara, N. S. Contractor, W. Hall, J. A. Hendler, and N. Shadbolt, "Web science: understanding the emergence of macro-level features on the world wide web." Foundations and Trends in Web Science, vol. 4, no. 2-3, 2013, 103–267.

[24] S. Lawrence, "Context in web search." IEEE Data Eng.Bull., vol. 23, no. 3, 2000, 25–32.

[25] L. Finkelstein, et al., "Placing search in context: The concept revisited." In Proceedings of the 10th international conference on World Wide Web, pages 406–414. ACM, 2001. [27] S. Lawrence. Context in web search. IEEE Data Eng. Bull., vol. 23, no. 3, 2000, 25–32.

[26] D. Kelly and J. Teevan, "Implicit feedback for inferring user preference: a bibliography." In ACM SIGIR Forum, volume 37, 2003, pages 18–28. ACM.

[27] T. Joachims, L. Granka, B. Pan, H. Hembrooke, and G. Gay, "Accurately interpreting clickthrough data as implicit feedback." In Proceedings of the 28th annual international ACM SIGIR conference on Research and development in information retrieval, 2005, pages 154–161. ACM.

[28] E. Agichtein, E. Brill, and S. Dumais, "Improving web search ranking by incorporating user behavior information." In Proceedings of the 29th annual international ACM SIGIR conference on Research and development in information retrieval, 2006, pages 19–26. ACM.

[29] Z. Dou, R. Song, and J.-R. Wen, "A large-scale evaluation and analysis of personalized search strategies." In Proceedings of the 16th international conference on World Wide Web, 2007, pages 581–590. ACM.

[30] G. Adomavicius and A. Tuzhilin, "Context-aware recommender systems." In Recommender systems handbook, 2011, pages 217–253. Springer.

[31] X. Shen, B. Tan, and C. Zhai, "Implicit user modeling for personalized search." In Proceedings of the 14th ACM international conference on Information and knowledge management, 2005, pages 824–831. ACM.

[32] J. Teevan, S. T. Dumais, and E. Horvitz, "Personalizing search via automated analysis of interests and activities." In Proceedings of the 28th annual international ACM SIGIR conference on Research and development in information retrieval, 2005, pages 449–456. ACM.

[33] D. Carmel, et al., "Personalized social search based on the user's social network." In Proceedings of the 18th ACM conference on Information and knowledge management, 2009, pages 1227–1236. ACM.

[34] A. Hannak, P. Sapiezynski, A. Molavi Kakhki, Krishnamurthy, D. Lazer, A. Mislove, and Wilson, "Measuring personalization of web search." In Proceedings of the 22nd international conference on World Wide Web, 2013, pages 527–538. International World Wide Web Conferences Steering Committee.

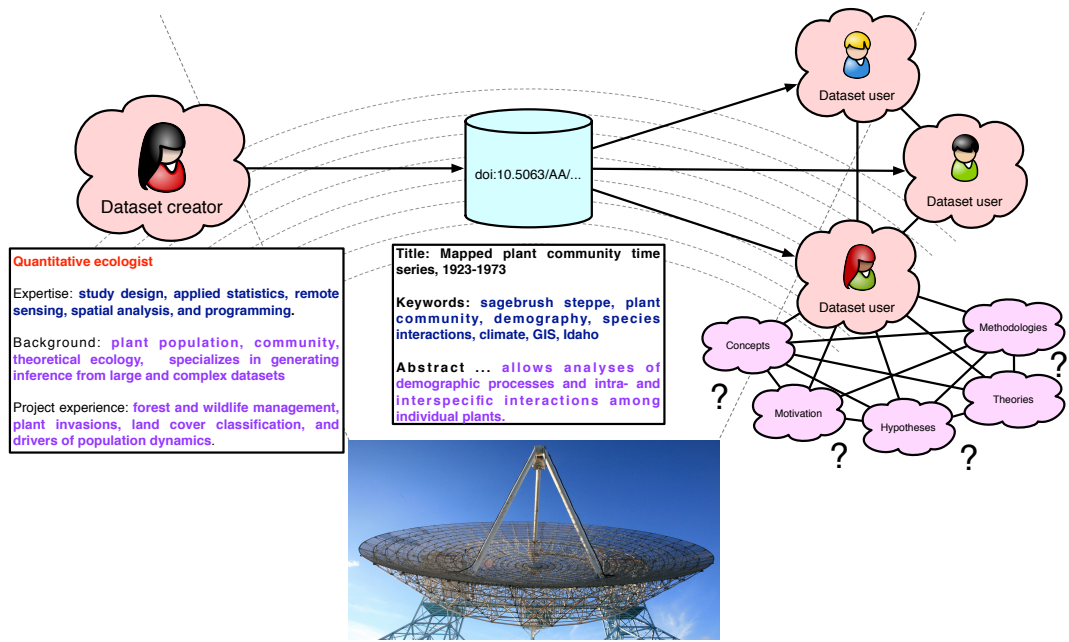

Figure 2. Geographic information observatories study the interactions and connections within a cyber community. We can observe unstructured data sources such as researcher websites, scientific articles and abstracts. Instrumenting the research community will allow us to get at deeper relationships. The observed relations in the nexus can be mined to discover the purposes for which data are fit.